\documentclass[11pt]{article}
\usepackage{amsmath,amssymb}
\usepackage{graphicx}
\newcommand{\beq}{\begin{equation}}
\newcommand{\eeq}{\end{equation}}
\newcommand{\ba}{\begin{array}}
\newcommand{\ea}{\end{array}}
\newcommand{\bea}{\begin{eqnarray}}
\newcommand{\eea}{\end{eqnarray}}
\newcommand{\bean}{\begin{eqnarray*}}
\newcommand{\eean}{\end{eqnarray*}}
\newtheorem{theorem}{Theorem}[section]
\newtheorem{prop}[theorem]{Proposition}
\newtheorem{lem}[theorem]{Lemma}

\newtheorem{defi}[theorem]{Definition}
\newtheorem{remark}[theorem]{Remark}
\newenvironment{rem}{\begin{remark} \rm}{\end{remark}}
\newcommand{\CH}{{\cal H}}

\newcommand{\CZ}{{\cal Z}}
\newcommand{\CS}{{\cal S}}

\newcommand{\CN}{{\cal N}}

\newcommand{\CM}{{\cal M}}
\newcommand{\CQ}{{\cal Q}}

\makeatletter
\@addtoreset{equation}{section}

\makeatother

\newcommand{\VV}{{\mathsf V}}

\newcommand{\FF}{{\mathsf F}}

\def\la{\lambda}

\newcommand{\cmp}[3]{Commun. Math. Phys. {\bf #1} (#2), #3}

\newcommand{\pla}[3]{Phys. Lett. {\bf A #1} (#2), #3}

\newcommand{\faa}[3]{Funct. Anal. Appl. {\bf #1} (#2), #3}

\newcommand{\lmp}[3]{Lett. Math. Phys. {\bf #1} (#2), #3}

\newcommand{\rmp}[3]{Rev. Math. Phys. {\bf #1} (#2), #3}

\newcommand{\jmp}[3]{J. Math. Phys. {\bf #1} (#2), #3}

\newcommand{\rref}[1]{(\ref{#1})} 
\def\mat2#1#2#3#4{{\left[\begin{array}{cc}
#1 & #2\\ #3 & #4 \end{array}\right]}}
\def\vec2#1#2{{\left[\begin{array}{c}
#1 \\ #2 \end{array}\right]}}
\newcommand{\del}{{\partial}}

\def\dsl#1{{\displaystyle #1}}

\def\parpu{\{\cdot,\cdot\}}
\def\dpt#1#2{\frac{\partial #1}{\partial t_{#2}}}

\def\Bdpt#1#2{\displaystyle{\frac{\partial #1}{\partial t_{#2}}}}
\def\H#1{H^{(#1)}}

\def\endpf{\par\hfill$\square$\par\medskip\par\noindent}
\def\hie{hierarch}
\def\vefi{vector field}
\def\man{manifold}
\def\bih{bi-Hamiltonian}
\def\bihm{\bih\ \man}
\def\MA{{\CM_{A}}}
\def\ham{Hamiltonian}
\def\restr{restriction}
\def\syml{symplectic leaf}
\def\symls{symplectic leaves}
\def\poib{Poisson bracket}
\def\poit{Poisson tensor}
\def\poip{Poisson pencil}
\def\poim{Poisson manifold}
\def\wrt{with respect to}
\def\kdv#1{${\mbox{KdV}}_{#1}$}

\def\alg{{\mathfrak g}}

\newcommand{\fraksl}{{\mathfrak{sl}}}
\newcommand{\CHp}{{H_+}}

\def\Tr{\mathrm{Tr}}
\def\V#1{{\VV^{(#1)}}}
\begin{document}
\begin{titlepage}
\null\vspace{0.5truecm}
\begin{center}
{\Huge A Bi-Hamiltonian Theory for \\
\vspace{.3truecm} Stationary KdV Flows
and their Separability}
\end{center}
\vspace{0.5truecm}
\makeatletter
\begin{center}
{\large
Gregorio Falqui${}^1$,
Franco Magri${}^2$,\\ \vspace{0.3truecm}
Marco Pedroni${}^3$
and Jorge P. Zubelli${}^4$}\\
\vspace{0.2truecm}
${}^1$ SISSA, Via Beirut 2/4, I-34014 Trieste, Italy\\
E--mail: falqui@sissa.it\\\vspace{0.3truecm}
${}^2$ Dipartimento di Matematica e Applicazioni\\
Universit\`a di Milano--Bicocca\\
Via degli Arcimboldi 8, I-20126 Milano, Italy\\
E--mail: magri@vmimat.mat.unimi.it\\\vspace{0.3truecm}
${}^3$ Dipartimento di Matematica, Universit\`a di Genova\\
Via Dodecaneso 35, I-16146 Genova, Italy\\
E--mail: pedroni@dima.unige.it\\ \vspace{0.3truecm}
${}^4$ IMPA \\
Est. D. Castorina 110, Rio de Janeiro,
RJ 22460, Brazil\\
E--mail: zubelli@impa.br
\end{center}
\makeatother
\vspace{.5truecm}
\begin{abstract}\noindent
We present a fairly new and comprehensive approach to the study of
stationary flows of the Korteweg-de Vries hierarchy.
They are obtained by means of a double
restriction process from a dynamical system in an infinite number of variables.
This process naturally provides us with a Lax representation of the flows, which
is used to find their bi-Hamiltonian formulation.
Then we prove the separability of these flows making use of their bi-Hamiltonian
structure, and we show that the variables of separation are supplied by the
Poisson pair.
\end{abstract}
\vspace{1.truecm}
\begin{center}
\end{center}
\vspace{1.truecm}
\end{titlepage}
\setcounter{footnote}{0}
\section{Introduction}

In this paper we present a comprehensive approach to the 
stationary
flows of the Korteweg--de Vries (KdV) equation. 
This is a quite classical subject,
and lies at the very heart of the modern theory of integrable systems in
finite and infinite dimensions. Its formulation
can be traced back at least
to~\cite{BoNo,PL75}. Then, together with its generalizations,
it has been subject of intensive studies.
See, for 
example, \cite{BuEnLe,DikBook,DKN,DMN,Kr77,Mum85}
and references therein. 
It was realized that such systems were among the prototypes of the
class of Algebraically Completely Integrable Hamiltonian Systems
\cite{AvM,MvM},
and, in particular, it has been proven that the classical Jacobi
formulas for projective embeddings of  hyperelliptic
Jacobian varieties found a very natural realization in such problems.

Hamiltonian aspects of those flows were
studied in a number of papers. To cite a few \cite{BoNo,DMN,PL75}. 
Two main approaches emerged.
In the first one, a Hamiltonian structure was given to the stationary
KdV flows,  looking at their variational properties~\cite{BoNo}.
Namely, such flows were regarded as
classical Euler--Lagrange equations associated to suitable Lagrangian and
Hamiltonian densities. In the second one (see, e.g.,~\cite{DKN}),
a set of suitable canonical coordinates was introduced on the stationary
manifolds,  somehow dictated by the algebro--geometrical structure of
the problem, and the fact that the flows were indeed Hamiltonian  was verified
at a later stage, via direct computation.  In the same way, a set
of action--angle variables was found.

After  the discovery of the \bih\ structure of the
$1+1$ dimensional KdV hierarchy,
the obvious problem of finding a corresponding bi-Hamiltonian structure for
its stationary flows was studied. A solution for this was
found~\cite{AFW,RW92,To95} with an ingenious 
study mainly based on the
recursion relations found by Alber~\cite{Alber}, the introduction of
special sets of coordinates, and the use of the Miura transformation.

The aim of this paper is to give a rather new and systematic perspective
to this circle of ideas,
focusing our attention  on the \bih\ aspects of the problem.
The next section is devoted to a quick description, 
in a simple example, 
of the stationary reductions of KdV
and to an illustration of the main
properties of these systems. Then,
we will present the plan and the important
points of the paper.

\section{A Preliminary View}
\label{sec:preview}

In this section we recall some known facts about the KdV \hie y
and its stationary reductions, and we present the problems to
be tackled in the next sections.

The KdV equation is the most famous example in the class of the so-called
integrable nonlinear PDEs.
It possesses a number of remarkable properties, in
particular:
\begin{enumerate}
\item It has an infinite sequence of integrals of motion;
\item It admits a Lax representation;
\item It is a \bih\ system;
\item The integrals of motion are the coefficients of a Casimir of the
Poisson pencil, so that the KdV equation can be seen as a 
{\em Gel'fand--Zakharevich system\/} (see below).
\end{enumerate}
There are of course relations between these properties. For example, the
conserved densities are the residues of the fractional powers of the Lax
operator. Also, they can be extracted from the \bih\ structure and
shown to commute with respect to both Poisson brackets. The associated
vector fields form the KdV \hie y, whose first members are
\begin{equation}
\label{KdVmem}
\begin{array}{c}
\dsl{\dpt{u}1=u_x},\qquad \dsl{\dpt{u}3=\frac14(u_{xxx}-6uu_x)}
\quad\mbox{(KdV equation)}      \vspace{2truemm}     \\
\dsl{\dpt{u}5=\frac1{16}(u_{xxxxx}-10uu_{xxx}-20 u_x u_{xx}+30 u^2 u_x)}\ .
\end{array}
\end{equation}
The KdV \hie y can be used to find
finite-dimensional reductions for the KdV equation, giving
rise to explicit solutions.
Indeed, the set of singular points of a (fixed) vector field of
the \hie y is a finite-dimensional manifold which is invariant
under the flows of the other vector fields,
due to the commutativity property. The (finite-dimensional) systems
obtained by restricting the KdV \hie y
to such invariant manifolds are called the {\em stationary reductions of
KdV\/}.

Let us 
consider explicitly the reduction ${\mbox{KdV}}_5$
corresponding to the third vector
field of the \hie y. The set of zeroes is given by
\begin{equation}
\label{KdV5con}
u_{xxxxx}-10uu_{xxx}-20 u_x u_{xx}+30 u^2 u_x=0\ ,
\end{equation}
and its dimension is 5, since we can
use the value of $u$, $u_x$, $u_{xx}$, $u_{xxx}$, and $u_{xxxx}$ at a
fixed point $x_0$ (i.e., the Cauchy data) as global coordinates.
For the sake of simplicity we put
\begin{equation}
u_0=u(x_0),\quad u_1=u_x(x_0),\quad u_2=u_{xx}(x_0),\quad
u_3=u_{xxx}(x_0),\quad  u_4=u_{xxxx}(x_0)\ .
\end{equation}
In order to compute the reduced equations of the first flow of
\rref{KdVmem}, we have to derive it with respect to $x$, and to use
the constraint \rref{KdV5con} and its differential consequences
to eliminate all the derivatives of order higher than 4. We
obtain the equations
\begin{equation}
\label{1kdv5}
\begin{array}{c}
\dsl{\dpt{u_0}1=u_1},\quad \dsl{\dpt{u_1}1=u_2},\quad
\dsl{\dpt{u_2}1=u_3},\quad \dsl{\dpt{u_3}1=u_4}, \vspace{2truemm} \\
\dsl{\dpt{u_4}1=10 u_0 u_3+20 u_1 u_2-30 {u_0}^2 u_1}\ .
\end{array}
\end{equation}
In the same way, for the KdV equation we get
\begin{equation}
\label{3kdv5}
\begin{array}{l}
\dsl{\dpt{u_0}3=\frac14(u_3-6u_0u_1)} \vspace{2truemm}\\
\dsl{\dpt{u_1}3=\frac14(u_4-6u_0u_2-6{u_1}^2)} \vspace{2truemm}\\
\dsl{\dpt{u_2}3=\frac14(4u_0u_3+2u_1u_2-30{u_0}^2u_1)} \vspace{2truemm}\\
\dsl{\dpt{u_3}3=\frac14(4u_0u_4+6u_1u_3+2{u_2}^2-30{u_0}^2u_2-
60u_0{u_1}^2)} \vspace{2truemm}\\
\dsl{\dpt{u_4}3=\frac14(10u_1u_4+10{u_0}^2u_3+10u_2u_3-100u_0u_1u_2
-60{u_1}^3-120{u_0}^3u_1)}
\end{array}
\end{equation}
As far as the restrictions of the other flows are concerned, it can
be shown that they are linear combination of \rref{1kdv5} and \rref{3kdv5}.

It is not surprising that the above mentioned properties of the
KdV equation hold also for the \kdv5 system. However, to the best of
our knowledge, it has not been made completely clear the way which
these properties pass from the KdV
\hie y to its stationary reductions, especially as far as the \bih\
structure is concerned. In any case, one can check that the
functions
\begin{equation}
\begin{array}{l}
H_0=\dsl{\frac1{16}(-2u_2u_4+6{u_0}^2u_4+{u_3}^2-12u_0u_1u_3
+16u_0{u_2}^2+12{u_1}^2u_2-60{u_0}^3u_2+36{u_0}^5)}
\vspace{2truemm} \\
H_1=\dsl{-\frac14(2u_0u_4-2u_1u_3+{u_2}^2-20{u_0}^2u_2+15{u_0}^4)}
\vspace{2truemm} \\
H_2=u_4-10 u_0u_2-5{u_1}^2+10 {u_0}^3
\end{array}
\end{equation}
are integrals of motion for \rref{1kdv5} and \rref{3kdv5}.
These systems have also a Lax formulation, i.e., they can be written as
\begin{equation}
\dpt{L}{i}=[A_i,L],\qquad i=1,3,
\end{equation}
where the Lax matrix $L$ depends on a parameter $\la$, and is given by
\begin{equation}
L=\frac1{16}\left(
\begin{array}{cc}
4u_1\la+u_3-6u_0 u_1 & 16\la^2 -8u_0\la+6{u_0}^2-2u_2
\vspace{2truemm}\\
\begin{array}{c} 16\la^3+8u_0\la^2+2\la(u_2-{u_0}^2)+ \vspace{-2truemm}\\
                 u_4-8u_0u_2-6{u_1}^2+6{u_0}^3\end{array}
& -4u_1\la-u_3+6u_0 u_1
\end{array}\right)\ .
\end{equation}
The matrices $A_i$ can be easily contructed from $L$ (see
Section~\ref{sec:lax}).

Finally, there are two compatible Poisson structures
giving a (bi)-Hamiltonian formulation of the \kdv5\ systems. The
corresponding Poisson tensors are
\begin{equation}
P_0=\left [\begin {array}{ccccc}
0&0&0&2&0\\
0&0&-2&0&-20u_0\\
0&2&0& 20u_0 &20 u_1\\
-2&0&-20u_0 &0& -140 {u_0}^2-20u_2\\
0&20u_0&-20u_1&140 {u_0}^2+20u_2 &0
\end {array}\right ]\nonumber
\end{equation}
and
\begin{equation}
P_1=\left [\begin {array}{ccccc}
0&\frac12&0&3u_0&6u_1\\
-\frac12&0&-3u_0&-3u_1&-4u_2-15{u_0}^{2}\\
0&3u_0&0&u_2+15{u_0}^{2} &u_3+30u_0u_1\\
-3u_0&3u_1&-u_2-15{u_0}^{2}&0&
                  \begin{array}{cc}u_4-40u_0u_2+\\
                   30{u_1}^{2}-60{u_0}^{3}\end{array}\\
-6u_1&4u_2+15{u_0}^{2}&-u_3-30u_0u_1 &
                  \begin{array}{cc}-u_4+40u_0u_2-\\
                   30{u_1}^{2}+60{u_0}^{3}\end{array}
                   &0\end {array}
\right ]\ .\nonumber
\end{equation}
If we call $X_1$ and $X_3$ the \vefi s of \kdv5, then the following
relations hold:
\begin{equation}
\begin{array}{c}
P_0dH_2 = 0\\
X_1 = P_0dH_1 = P_1dH_2\\
X_3 = P_0dH_0 = P_1dH_1\\
P_1dH_0 = 0\ .
\end{array}
\end{equation}
They can be collected in the statement that the function
$H(\la):=H_2\la^2+H_1\la+H_0$ is a Casimir of the {\em Poisson
pencil\/} $P_\la:=P_1-\la P_0$, that is,
\begin{equation}
P_\la dH(\la)=0\ .
\end{equation}
Therefore, $X_1$ and $X_3$ are the \bih\ \vefi s associated with a
polynomial Casimir of a \poip\ of maximal rank. In a word, they are 
Gel'fand--Zakharevich (GZ) systems \cite{GZ93}.

The importance of the stationary reductions of the KdV \hie y
(and, more generally, of the stationary reductions of the
Gel'fand--Dickey \hie ies) lies in the fact that the reduced
equations can be solved by means of the classical method
of separation of variables. This was noticed in the early works on the subject.
It is also known how to construct the variables of separation starting from
the Lax matrix. We will show that the separability
of these systems is a particular instance of a general result,
which is valid for quite a wide class of \bihm s.

In Section \ref{sec:stared} we give a rather unconventional presentation of the
stationary reductions of KdV. Our priviledged  starting point  is a picture of
the KP hierarchy as a system of ordinary differential equations, called the {\em
Central System} (CS) in~\cite{cfmp5,fmp1}.
Starting from there, by means of a double reduction
process we can describe quite explicitly the stationary reductions of KdV, and,
in Section \ref{sec:lax}, we are able to give a Lax representation of these
systems, with a Lax matrix depending polynomially on a parameter.
This representation is used to show that the flows are \bih. This is done in two
steps. First, in Section \ref{sec:bihlax} we recall the \bih\ structure on
matrix polynomials and show that the \ham\ \vefi s (\wrt\ the \poip) admit
a Lax formulation. This property is conserved after a suitable \bih\ reduction
process. Then, in Section \ref{sec:bihstr}, we identify the phase space of
the stationary reductions of KdV with a reduced \bihm, and we show that they are
GZ systems. In Section \ref{sec:sepbih} we state
(referring to \cite{fmp2} for a more detailed discussion) a theorem ensuring
that, under some additional assumptions, the GZ systems are separable in
coordinates that are naturally associated with the \bih\ structure. Finally, in
Section \ref{sec:sepsta} we show that this theorem can be applied to the
stationary reductions of KdV, and that the variables of separation can be
constructed algebraically.

Summing up, in this paper we present a somewhat self--contained
approach to the study of stationary flows of KdV, and we use them as a
laboratory to test ideas of the \bih\ geometry, from the GZ theory to
the separation of variables. 
In our opinion, such a set up provides a
comprehensive formulation of results which, although for the most
part already available in the literature,   
would perhaps acquire a deeper meaning under this perspective.

\section{KdV Stationary Reductions}
\label{sec:stared}

In this section we give a self-contained presentation of the
stationary reductions of the KdV \hie y, using the formalism
developed in~\cite{cfmp5,fmp1} for the KP theory. Our starting point
is the Central System (CS), a family of dynamical systems with
${\Bbb N}\times{\Bbb N}$ degrees of freedom. A first (stationary)
reduction gives rise to the ${\mbox{CS}}_2$ \hie y, with $\Bbb N$
degrees of freedom. Then a further restriction leads to
finite-dimensional systems that coincide with the stationary reductions
of KdV.

We consider the space $\CH$ of sequences $\{\H{k}\}_{k\ge 1}$ of
Laurent series
having the form $H^{(k)}=z^k+\sum_{l\ge 1}H^k_l z^{-l}$,
where  $H^k_l$ are (complex) scalars that play
the role of coordinates on $\CH$.
On such phase space $\CH$
we define a family of vector fields
as follows. We associate with a point $\{\H{k}\}_{k\ge 1}$ in $\CH$
the linear span $\CHp=\langle \H{0}, \H{1},\H{2},\ldots \rangle$,
where $\H{0}=1$.
The {\em defining equation} for the $j$--th vector field $X_j$ of the
family,
to be referred to as the {\em Central System} (CS), is the invariance relation
\begin{equation}
\left(\dpt{}{j}+\H{j}\right)\CHp\subset\CHp\ .
\end{equation}
This relation is equivalent to the (explicit) equations
\begin{equation}
\label{CS}
\frac{\partial H^{(k)}}{\partial t_j}= -H^{(j)}H^{(k)}+ H^{(j+k)}
+\sum_{l=1}^kH^j_lH^{(k-l)}+\sum_{l=1}^jH^k_lH^{(j-l)}, \qquad k\ge 1.
\end{equation}
\begin{rem}
From \rref{CS} it is evident that the exactness property
\begin{equation}
\label{exapro}
\dsl{\dpt{}{k}\H{j}=\dpt{}{j}\H{k}}
\end{equation}
holds. Moreover, it can be shown \cite{cfmp5} that the flows of
the CS commute.
\end{rem}

\begin{rem}
There is a very tight relation between the CS and the linear flows on the Sato
Grassmannian~\cite{SS,SW85,DJKM}. This relation is discussed in~\cite{fmp1}, 
where
the classical result of Sato on the linearization of the KP hierarchy is
recovered from the point of view of the \bih\ geometry.
\end{rem}

Since the CS is a family of commuting \vefi s, we can reduce it in many
different ways. By means of a suitable combination of such reduction processes,
the so-called fractional KdV hierarchies were obtained in~\cite{cfmp5}. Now we
will show how the stationary reductions of KdV can be derived from the CS.
The commutativity of the flows implies that the
set $\CZ_2$ of zeroes of the vector field $X_2$,
defined by the quadratic equations
\begin{equation}\label{zerixn}
H^{(k+2)}-H^{(k)}H^{(2)}+\sum_{l=1}^kH^2_lH^{(k-l)}+H^k_1 H^{(1)}+H^k_2=0,
\end{equation}
is an invariant submanifold for CS.
Moreover, on $\CZ_2$ we have
\begin{equation}\label{0cu}
\dpt{\H{2}}{j}=\dpt{\H{j}}{2}=0,
\end{equation}
due to the exactness property \rref{exapro}.
Therefore, the manifold $\CZ_2$ is foliated by invariant
submanifolds defined by the equation $\H{2}=\mbox{constant}$.
Among all these leaves, the submanifold $\CS_2$
defined by the simple constraint
\begin{equation}
\H{2}=z^2 \;\label{snc}
\end{equation}
is particularly relevant.
At the points of $\CS_2$ equation \rref{zerixn} takes the form
\begin{equation}\label{constr}
\H{k+2}=z^2\H{k}-H^k_1\H{1}-H^k_2 \;,
\end{equation}
and allows us to recursively compute the Laurent
coefficients of $\H{k}$, for $k>2$,
in terms of the coefficients of 
$h:=\H{1}$.
Hence, $\CS_2$ is parametrized by the coefficients $\{h_l\}_{l\ge 1}$ of
$h$. Equation \rref{constr} also shows that $z^2(H_+)\subset H_+$,
so that on $\CS_2$ the elements $\{z^{2j},z^{2j} h\}_{j\ge 0}$ form
a basis in $H_+$. Thus, we have that
\begin{equation}
\label{CS2cur}
\H{k}=p_k(z^2)+q_k(z^2)h\qquad\mbox{on $\CS_2$},
\end{equation}
where $p_k$ and $q_k$ are polynomials. This can also be seen directly
from equation \rref{constr}. Moreover,
there is only one Laurent series of
the previous form satisfying the asymptotic condition
$\H{k}=z^k+O(z^{-1})$ as
$z\to\infty$.
\begin{defi} {\bf (see \cite{cfmp5})}. The restriction of CS to the
invariant submanifold $\CS_2$ is called the ${\mbox CS}_2$ \hie y.
\end{defi}

The restricted vector fields are given by
\begin{equation}
\label{CS2}
\frac{\partial h}{\partial t_j}= -h H^{(j)}+ H^{(j+1)}
+\sum_{l=1}^j h_l H^{(j-l)}+H^j_1, \qquad j\ge 1,
\end{equation}
where the $\H{j}$ must be written in terms of $h$ according to
\rref{CS2cur}. If we denote with $H_-$ the span of the negative powers
of $z$, and with $\pi_-$ the projection on $H_-$ according
to the decomposition $H_+\oplus H_-$, then the equations \rref{CS2}
can be written in the more compact form
\begin{equation}
\label{CS2bis}
\frac{\partial h}{\partial t_j}= -\pi_-\left(q_j h^2\right)\ .
\end{equation}
Notice that $\H{2k}=z^{2k}$, so that \rref{exapro} implies that
the even flows of ${\mbox{CS}}_2$ are trivial.

The finite--dimensional systems
that are the main subject of this paper are those  obtained
by restricting
the ${\mbox{CS}}_2$ flows to the manifold 
of zeroes of the
$(2g+1)$-st vector field of ${\mbox{CS}}_2$.  We will call such systems the
${\mbox{KdV}}_{2g+1}$ systems, since they are (equivalent to) the stationary
reductions of the KdV \hie y, as we are going to show at the end of this
section.
The constraint which defines the phase space $\CM_{2g+1}$
of the ${\mbox{KdV}}_{2g+1}$ system is
\begin{equation}
\label{statcons}
\frac{\partial h}{\partial t_{2g+1}}=-\pi_-\left(q_{2g+1} h^2\right)
=0\ .
\end{equation}
A direct inspection shows that
this constraint gives all the coefficients of $h$ in terms of the first
$(2g+1)$, i.e., $h_1,\dots,h_{2g+1}$. In other words,
the dimension of the phase space of
the ${\mbox{KdV}}_{2g+1}$ system equals $2g+1$.
The equations are given by the first $2g+1$ components of \rref{CS2bis},
after substituting the constraints \rref{statcons}. In the case of \kdv5\ there
are two independent vector fields:
\begin{equation}
\label{eqkdv5}
\begin{array}{ll}
\Bdpt{h_1}1=-2h_2 & \Bdpt{h_1}3= -2h_4+2h_1h_2 \vspace{2truemm}   \\
\Bdpt{h_2}1=-2h_3-{h_1}^{2} & \Bdpt{h_2}3= -2h_5+{h_2}^{2}+{h_1}^{3}
\vspace{2truemm}   \\
\Bdpt{h_3}1=-2h_1h_2-2h_4 & \Bdpt{h_3}3= -2h_1h_4+4{h_1}^{2}h_2-2h_3h_2    
\vspace{2truemm} \\
\Bdpt{h_4}1=-2h_5-{h_2}^{2}-2h_1h_3 & \Bdpt{h_4}3=-2{h_3}^{2}-2h_2h_4
+2h_1{h_2}^{2}+{h_1}^{4}+{h_1}^{2}h_3  \vspace{2truemm}  \\
\Bdpt{h_5}1=-4h_3h_2+2{h_1}^{2}h_2-4h_1h_4 &
\Bdpt{h_5}3=2{h_1}^{2}h_4-4h_3h_4+2{h_1}^{3}h_2
\end{array}
\end{equation}
These are, up to the coordinate change \rref{utohj}, the equations
\rref{1kdv5} and \rref{3kdv5}.

We remark that along the flows of \kdv{2g+1}\ the relations \rref{exapro}
take the form
\begin{equation}
\dpt{\H{2g+1}}{j}=0\ ,
\end{equation}
showing that all the coefficients of $\H{2g+1}$ are integrals of
motion. Therefore our presentation of the KdV stationary reductions
carries directly the conserved quantities of the flows.
Moreover, in the next section we will show that the Lax
representation also arises in a natural way. We end this section with
the following: 
\begin{rem}
The usual KdV hierarchy in $1+1$ dimensions is described in~\cite{cfmp5}
as a projection of ${\mbox{CS}}_2$ along the integral curves
of the first \vefi\ of the \hie y,
\[
\dpt{h}1=-h^2+z^2+2h_1\ .
\]
Indeed, if we put $x=t_1$ and $u=2h_1$, then the previous equation
takes the form $h_x+{h}^2=u+z^2$ and allows us to write the $h_j$
as polynomials in $u$ and its $x$-derivatives:
\begin{equation}
\label{utohj}
\begin{array}{l}
h_1=\frac12u\\
h_2=-\frac14u_x\\
h_3=\frac18(u_{xx}-u^2)\\
h_4=-\frac1{16}(u_{xxx}-4uu_x)\\
h_5=\frac1{32}(u_{xxxx}-6uu_{xx}-5u_x^2+2u^3)\\
\ \vdots
\end{array}
\end{equation}
Thus equations \rref{CS2} become partial differential
equations for the variable $u$, and are the KdV \hie y. But we can
also use the system \rref{utohj} to recover ${\mbox{CS}}_2$ from
the KdV \hie y, so that we can pass back and forth
from one \hie y to the other.
This shows that the \kdv{2g+1} systems that we have
introduced coincide with the usual stationary reductions of KdV.
The first $(2g+1)$ equations of the system \rref{utohj} represent
the change between our coordinates $(h_1,\dots,h_{2g+1})$ and
the ones usually considered in the literature, namely $(u,u_x,
\dots,u^{(2g+1)})$.
\end{rem}

\section{The Lax Representation}
\label{sec:lax}

In this section we show that there is a quite natural (Zakharov-Shabat)
zero-curvature representation for the ${\mbox{CS}}_2$ system, entailing
a Lax representation for the ${\mbox{KdV}}_{2g+1}$ \hie y.

We know from the previous section that in the ${\mbox{CS}}_2$ theory
every element in $H_+$ can be written as a
linear combination of 1 and $h$ with coefficients that are polynomials in
$\la:=z^2$. Then to each point of the manifold
$\CS_2$ (that is, to each series $h=z+\sum_{l\ge1} h_l z^{-l}$) we can
associate a  family of $2\times 2$ matrices
$\V{j}(\lambda)$
depending {\em polynomially} on $\lambda$,
defined by the relation
\begin{equation}
\label{eq:e1}
\left(\dpt{}{j}+\H{j}\right) \vec2{1}{h}=\V{j}\vec2{1}{h}.
\end{equation}
Since the even flows are trivial,
we will be interested only in the matrices of odd index.
The first three of them are given by
\begin{eqnarray*}
&\V{1}=
\left [\begin {array}{cc} 0&1
\\ \lambda+2\,h_{{1}}&0\end {array}\right ]\qquad
 \V{3}=\left [\begin {array}{cc} -h_{{2}}&\lambda-h_{{1}}
\\{\lambda}^{2}+h_{{1}}\lambda+2h_{{3}}-{h_1}^2 &h_2
\end{array} \right]\\
&\V{5}=
\left [\begin {array}{cc} -h_{2}\lambda-h_4+h_1h_2&{\lambda
}^{2}-h_{{1}}\lambda-h_3+{h_1}^2\\
\begin{array}{c}
{\lambda}^{3}+h_{{1}}{\lambda}^{2}+h_{{3}}\lambda+\vspace{-2truemm}\\
2h_5-2h_1h_3-{h_2}^{2}+{h_1}^3 \end{array}
&h_{{2}}\lambda-h_{{1}} h_{{2}}+h_{{4}}\end {array}\right ]\ .
\end{eqnarray*}
The commutativity of the flows and the ``abelian''
zero-curvature relation \rref{exapro}
imply that
\begin{equation}
\label{preZS}
\left(\dpt{}{j}\V{k}-\dpt{}{k}\V{j}+\left[\V{k},\V{j}\right]
\right) \vec2{1}{h}=0.
\end{equation}
Since the entries of the matrix appearing in the previous equation
are polynomials in $\la$, and the elements $\{\la^{j},\la^{j} h\}_{j\ge 0}$
are linearly independent in $H_+$, it follows that the
zero--curvature relations
\begin{equation}\label{ZS}
\dpt{}{j}\V{k}-\dpt{}{k}\V{j}+\left[\V{k},\V{j}\right]=0
\end{equation}
hold.

If we restrict to the set $\CM_{2g+1}$ of the stationary points of
the $(2g+1)$--st vector field of ${\mbox{CS}}_2$,
the zero--curvature representation naturally gives rise to Lax
equations for the matrix $\V{2g+1}$,
\begin{equation}
\label{eq:Lax}
\dpt{}{k}\V{2g+1}=\left[\V{k},\V{2g+1}\right].
\end{equation}
The following proposition will be useful in Section \ref{sec:bihstr}, and
shows that these Lax equations faithfully represent
the \kdv{2g+1} system.

\begin{prop}
The matrices $\V{2k+1}$ of the ${\mbox{CS}}_2$ \hie y have the
following properties:
\begin{enumerate}
\item The matrix $\V{2k+1}$ depends only on $(h_1,\dots,h_{2k+1})$,
and the map $(h_1,\dots,h_{2k+1})\mapsto\V{2k+1}$ is injective;
\item The trace of $\V{2k+1}$ is zero;
\item For $i\le k$ one has
\begin{equation}
\label{forV}
\V{2i+1}=(\la^{i-k}\V{2k+1})_+ -\alpha_{ik}
\left[\begin{array}{cc} 0 & 0\\ 1 & 0\end{array}\right],
\end{equation}
where $(\cdot)_+$ denotes the projection on the nonnegative powers
of $\la$, and $\alpha_{ik}$ is the entry $(1,2)$ of the coefficient
of $\la^{k-i-1}$ in $\V{2k+1}$.
\end{enumerate}
\end{prop}
{\bf Proof.} First of all we observe that, almost by definition,
\begin{equation}
\label{pqtoV}
\V{2k+1}=
\left[\begin{array}{cc} p_{2k+1} & q_{2k+1}\\
\begin{array}{c}\la^{k+1}+\sum_{l=0}^{k}h_{2l+1}\la^{k-l}+\vspace{-2truemm}\\
           \sum_{l=1}^k h_{2l} p_{2k-2l+1}+H_1^{2k+1} \end{array} &
\sum_{l=1}^{k}h_{2l} q_{2k-2l+1}\end{array}\right]\ .
\end{equation}
Then, we notice that equation \rref{constr}
implies the recursion formulas
\begin{equation}
p_{2k+1}=\la p_{2k_1}-H^{2k-1}_2\ , \qquad
q_{2k+1}=\la q_{2k_1}-H^{2k-1}_1\ .
\end{equation}
Thus, by induction, we obtain
\begin{equation}
\label{pqfor}
p_{2k+1}=-\sum_{l=1}^k H^{2l-1}_2 \la^{k-l}\ ,\qquad
q_{2k+1}=\la^k-\sum_{l=1}^k H^{2l-1}_1 \la^{k-l}\ .
\end{equation}
In order to express the coefficients $H_2^{2l-1}$ and
$H_1^{2l-1}$ in terms of the $h_l$, we use the identity
\[
H^{2k-1}_{l}=H^{1}_{2k+l-2}-\sum_{i=1}^{k+l-3}H^{1}_{l+2i-2}H^{2k-2i-1}_{1},
\]
which can be  proved by induction on $k$ using again~\rref{constr}.
In particular, we have

\begin{eqnarray}
H^{2k-1}_{1}&=&H^{1}_{2k-1}-\sum_{i=1}^{k-2}H^{1}_{2i-1}H^{2k-2i-1}_{1}\\
H^{2k-1}_{2}&=&H^{1}_{2k}-\sum_{i=1}^{k-1}H^{1}_{2i}H^{2k-2i-1}_{1}
\ .\label{CS2con2}
\end{eqnarray}
This allows us to control the appearance of the $h_i=H_i^1$ in
$p_{2l+1}$ and $q_{2l+1}$, and leads to the proof of the first
assertion.

The second statement tantamounts to  
$p_{2k+1}+\sum_{l=1}^{k}h_{2l} q_{2k-2l+1}=0$, and is easily proved
by inserting \rref{pqfor} and using \rref{CS2con2}.

As far as the last assertion is concerned, we use the following
consequences of \rref{pqfor}:
\begin{equation}
(\la^{i-k} p_{2k+1})_+=p_{2i+1},\qquad
(\la^{i-k} q_{2k+1})_+=q_{2i+1}\ .
\end{equation}
This gives, using \rref{pqtoV},
\begin{equation}
(\la^{i-k}\V{2k+1})_+=\V{2i+1}-H_1^{2i+1}
\left[\begin{array}{cc} 0 & 0\\ 1 & 0\end{array}\right]\ .
\end{equation}
Since $-H_1^{2i+1}$ is the coefficient of $\la^{k-i-1}$ in
$q_{2k+1}$, equation \rref{pqtoV} shows that we are done.
\endpf

So we have seen that the double reduction process of the Central System
outlined in Section \ref{sec:stared} provides us with a natural Lax
representation of the (commuting) vector fields
of the ${\mbox{KdV}}_{2g+1}$ system. Actually, as it was explained
in~\cite{fmp1}, the Central System can be seen as an outgrowth of the \bih\
properties of the KdV \hie y. It is thus natural to look for a \bih\ structure
of the ${\mbox{KdV}}_{2g+1}$ system. Unfortunately, we are not in a
position to derive
such a property from the Central System itself,
but rather we have to rely on the Lax representation discussed so far.
Namely, in the next two sections we
will establish the \bih\ nature of ${\mbox{KdV}}_{2g+1}$, showing that
it comes from the
general theory of \bih\ systems defined on matrices depending
polynomially on a parameter.

\section{Lax Equations and bi-Hamiltonian Systems}
\label{sec:bihlax}

In the previous section we have associated with every point of the phase space
$\CM_{2g+1}$ of ${\mbox{KdV}}_{2g+1}$ a Lax matrix $\V{2g+1}$, and we have seen
that this matrix gives a Lax representation of the flows. To give these flows
a \bih\ formulation, we will address in this section a general problem,
concerning the relation between Lax matrices and \bih\ structures. We will
describe a class of \bihm s whose (bi-)\ham\ flows have a Lax formulation, and
show that this formulation survives a reduction process of Marsden-Ratiu
type.  
Since $\V{2g+1}$ depends polynomially on $\la$, it is quite
natural to consider the multi-Hamiltonian structures  defined on
$\frak g$--valued polynomials (see~\cite{RSTS,MaMag}),
where $\frak g$ is a Lie algebra of
matrices such that the trace of the product is nondegenerate.
More precisely, for a  fixed matrix $A\in \alg$,
let us consider the space
\begin{equation}
\label{matAg}
\MA:=\{X(\lambda)=\lambda^{n+1} A+\sum_{i=0}^{n}\lambda^i X_i
\mid X_i\in\alg\}\ ,
\end{equation}
which is clearly in a 1-1 correspondence with the space $\bigoplus_{i=
0}^{n} \frak g$ of $(n+1)$--tuples of matrices in $\frak g$.
The tangent and the cotangent space at a point
of $\MA$ can also be identified with $\bigoplus_{i=0}^{n} \frak g$,
using the pairing 
\begin{equation}
\langle (V_0,\dots,V_n),(W_0,\dots,W_n)\rangle
=\sum_{i=0}^n \Tr(V_i W_i)\ .
\end{equation}
If $F$ is a function on $\MA$, we will denote its differential
by
\[
dF=\left(\frac{\del F}{\del X_0},\dots,\frac{\del F}{\del X_n}\right)\ .
\]
It is known that on $\MA$ there is an $(n+2)$--dimensional web of
(compatible) \poib s, and that this web is associated with a family of classical
$R$--matrices. Nevertheless, it turns out that in our case the
relevant Poisson pair is given by the first two brackets of the
above mentioned family.
The first \poit,
as a map from the cotangent to the tangent space, is given by
\begin{equation}
\label{eqcp0}
P_{0}:
\left( \begin{array}{c} W_0 \\  W_1\\ \vdots \\ \vdots\\ W_{n-1}
\end{array} \right)
\mapsto
\left( \begin{array}{c} \dot X_0\\ \dot X_1\\ \vdots \\ \vdots \\ \dot
X_{n-1}\end{array} \right) =
\left( \begin{array}{ccccc}
\left[X_1, \cdot\right] &  \left[X_2, \cdot\right]&
\cdots  & \cdots &\left[A, \cdot\right] \\
\left[X_2, \cdot\right]&\cdots & \cdots &\left[A, \cdot\right]& 0\\
\vdots & \dots & \cdot &&\\
\vdots & \cdot &&&\\
\left[A, \cdot\right] & 0 &\cdots& & 0
\end{array}\right)
\left( \begin{array}{c} W_0 \\  W_1\\ \vdots \\ \vdots\\ W_{n-1}
\end{array} \right)
\mbox{ ,}
\end{equation}
while the second one is
\begin{equation}
\label{eqcp1}
P_{1}:
\left( \begin{array}{c} W_0 \\  W_1\\ \vdots \\ \vdots \\ W_{n-1}
\end{array} \right)
\mapsto
\left( \begin{array}{c} \dot X_0\\ \dot X_1\\ \vdots \\ \vdots \\
\dot X_{n-1}\end{array} \right) =
\left(\begin{array}{ccccc} -\left[X_0,
\cdot\right] &0& \cdots & \cdots  & 0\\
0 & \left[X_2, \cdot\right]& \left[X_3,\cdot\right] &
\cdots &\left[A, \cdot\right] \\
0 & \left[X_3, \cdot\right]& \cdots & \cdot &\\
\vdots & \vdots & \cdot && \\
0  &\left[A, \cdot\right]& \cdots & \cdots & 0\end{array}\right)
\left( \begin{array}{c} W_0 \\  W_1\\ \vdots \\ \vdots
\\ W_{n-1} \end{array} \right) \mbox{ .}
\end{equation}
The associated \poib s are given by
$\{F,G\}_i=\langle dF,P_i dG\rangle$, where $i=1,2$, and $F$, $G$ are
functions on $\MA$.
The \poit s \rref{eqcp0} and \rref{eqcp1} satisfy the remarkable property that
every linear
combination of them is still a \poit. For this reason one says that they are
{\em compatible}, and that $\MA$ is a \bihm.

Let us consider now the Poisson pencil $P_\la:=P_1-\la P_0$.
It is important to notice that its Hamiltonian \vefi s admit a Lax
representation, as shown in the following:
\begin{prop}
\label{bihtolax}
Let $F$ be a function on $\MA$, and ${\dsl{\frac{\del X(\la)}{\del
t_\la}}}=P_\la dF$ the
\ham\ \vefi\ associated by $P_\la$ to $F$. Then,
\begin{equation}
\frac{\del X(\la)}{\del t_\la}=\left[\frac{\del F}{\del X_0},X(\la)\right]\ .
\end{equation}
\end{prop}
{\bf Proof.} Use the expressions \rref{eqcp0} and \rref{eqcp1} to
compute the \vefi\ $P_\la dF$, then identify the parameter $\la$
appearing in the Poisson pencil with the $\la$ in $X(\la)$.
\endpf

At this point we want to enlarge the class of \bihm s giving rise to systems
admitting
a Lax representation, having also in mind the case of the stationary
reductions of KdV. To this aim, it is important to
recall a reduction theorem \cite{CMP} allowing us to ``move'' the
\bih\ structure from a given (``big'') \man\ to a smaller one. This
result is a particular case of a theorem by Marsden and Ratiu for
\poim s \cite{MR}, and can be applied to a general \bihm. Here, for
the sake of simplicity, we will describe it only in the case at hand.
The central point is that a Lax representation can be found also
for the \vefi s that are \ham\ \wrt\ the reduced Poisson pencil.

\begin{figure}
\centering
\includegraphics[width=.4\textwidth]{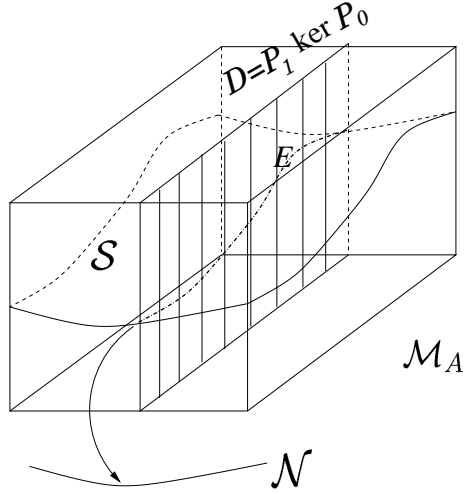}
\caption[The bi-Hamiltonian reduction]{ \label{bifig} The bi-Hamiltonian
reduction}
\end{figure}

The first step of the reduction process
is to fix a \syml\ $\CS$ of $P_0$. Then
we introduce the
distribution $D=P_1(\mbox{Ker}\, P_0)$, which is integrable thanks to the
compatibility between $P_0$ and $P_1$.
From the explicit form \rref{eqcp0}--\rref{eqcp1} of the \poit s,
it is easy to see that the vector fields in $D$ have the Lax form
$\dot X(\la)=[W_0,X(\la)]$, for a suitable $W_0$.
Let us denote by $E$ the intersection of $D$ with $T\CS$.
The statement of the \bih\ reduction theorem is that the quotient manifold
$\CN=\CS/E$ inherits from $\MA$ a \bih\ structure. In order to compute
the reduced Poisson bracket $\{f,g\}_\la$ 
between two functions $f$, $g$ on $\CN$, we
consider them as functions on $\CS$, invariant along the leaves of
$E$. Then, we choose functions $F$ and $G$ on $\MA$ which extend $f$
and $g$, and annihilate the distribution $D$. Their Poisson bracket
$\{F,G\}_\la$ is still invariant along $D$, and therefore defines 
a function on $\CN$, which is independent of the choice of the 
prolongations $F$ and $G$.

Let us consider now a given \ham\ \vefi\ $X_f$ (with respect to the Poisson
pencil) on $\CN$, with \ham\ $f$. If $F$ is a prolongation of $f$, the
\vefi\ $X_F:=P_\la dF$ is easily seen to be tangent to $\CS$ and to
project onto $X_f$.   
We are going to show that $X_f$ inherits a Lax representation from the one of
$X_F$.
To this aim, we suppose that there exists a submanifold $\CQ$ of
$\CS$, which is transversal to the distribution $E$. In other words,
$\CQ$ is the image of a section of the bundle $\pi:\CS\to\CN$.
Then $\CQ$ is
diffeomorphic to $\CN$, and inherits a \bih\ structure. The
representative of $X_f$ on $\CQ$ is simply found by decomposing
the restriction of $X_F$ to $\CS$ according to the splitting
$T\CS=T\CQ\oplus E$. Since we have seen that $E$ is spanned by \vefi s
having a Lax form, we have proved the following:
\begin{prop}
Let $\CQ\subset\CS$ be transversal to $E$. Then, the \vefi s on $\CN$
which are \ham\ with respect to the Poisson pencil admit a Lax
representation on $\CQ$.
\end{prop}
Therefore the \bih\ reduction implies, in this case, a reduction of
the Lax formulation.

\section{The \bih\ Structure of the KdV Stationary Reductions}
\label{sec:bihstr}

The aim of this section is to show that the ${\mbox{KdV}}_{2g+1}$
systems introduced in Section \ref{sec:stared} admit a \bih\
formulation. To do this, we are going to exploit the Lax representation found in
Section \ref{sec:lax} and the results of the preceding section.

The form of the Lax matrix $\V{2g+1}$ found in Section \ref{sec:stared}
suggests to choose $\alg=\fraksl(2)$, $n=g$, and
\[
A=\left[\begin{array}{cc} 0 & 0\\ 1 & 0\end{array}\right]\ .
\]
Therefore, the dimension of $\MA$ is $3(g+1)$.
The Lax matrix $\V{2g+1}$ defines an embedding of the
${\mbox{KdV}}_{2g+1}$ phase space into $\MA$. At this point two
natural questions arise:
\begin{enumerate}
\item Does this submanifold inherit from $\MA$ the \bih\
structure?
\item If so, are the \vefi s of ${\mbox{KdV}}_{2g+1}$ \bih\ with
respect to this structure?
\end{enumerate}
We will see that the answer to both questions is yes.

In order to answer the first question we need a careful
description of the \symls\ of $P_0$, as given by
\begin{lem}
\label{lem:syml}
The \symls\ of $P_0$ have dimension $2(g+1)$. Moreover,
let $H(\la):\MA\to \Bbb C$ be defined as $H(X(\la)):=\frac12\Tr X(\la)^2$
and let the $H_i$ be the coefficient of $\la^i$ in $H(\la)$.
Then, the functions
$H_{2g+1},\dots,H_{g+1}$ are functionally independent Casimirs of $P_0$.
Consequently, the symplectic leaves of $P_0$ are the level surfaces of the
previous Casimirs.
\end{lem}
{\bf Proof.} From \rref{eqcp0} the kernel of $P_0$ is easily seen to be given by
the covectors $\left[W_0,\dots,W_{g}\right]^T$ such that $W_i=\alpha_i A
+\sum_{l=1}^{i} \alpha_{i-l} X_{g+1-l}$, where the $\alpha_i$,
$i=0,\dots,g$, are
arbitrary. This shows that $\mbox{dim} (\mbox{Ker} P_0)=n$, so that
the dimension of the symplectic leaves is $2(g+1)$. In order to check that
$H_{2g+1},\dots,H_{g+1}$ are Casimirs, it is sufficient to verify that the
differential of $H_i$ is the 1-form
$\left[X_i,X_{i-1},\dots,X_{i-g}\right]^T$, where $X_k:=0$ if $k<0$.
\endpf

One can easily show that the \syml\ defined by $H_i=c_i$, with $g+1\le i\le
2g+1$, can be parametrized as
\begin{equation}\label{paraS}
X(\lambda)=\lambda^{g+1}A+\sum_{j=0}^{g}\lambda^j \mat2{p_{j}}{r_j}{q_{j}}{-
p_{j}}\ ,
\end{equation}
where $p_j$ and $q_j$ are free parameters, and $r_j$ is a function of
$(p_{j+1},q_{j+1},\dots,p_{g},q_{g})$ and the values $(c_{g+j+1},\ldots, 
c_{2g+1})$ of the Casimirs.

As far as the distribution $D=P_1(\mbox{Ker} P_0)$ is concerned,
in this case it is tangent to the
\symls\ of $P_0$. Indeed, from the explicit form \rref{eqcp0}--\rref{eqcp1}
of the \poit s
it is easy to see that $D$ is the 1-dimensional distribution spanned by the
vector field
\begin{equation}
\label{genD}
{\dot X}(\la)=[A,X(\la)]\ .
\end{equation}
This also shows that the integral leaves of
$D$ are simply the orbit of the action given by simultaneous
conjugation of the
isotropy subgroup of $A$, but we will never use this fact.

Now we are ready to endow the phase space $\CM_{2g+1}$ of
${\mbox{KdV}}_{2g+1}$ with the structure of a \bihm. This follows
from the fact that the map assigning to each point of $\CM_{2g+1}$
the corresponding Lax matrix $\V{2g+1}$ defines a submanifold
of a suitable \syml\ of $P_0$, which is transversal to the distribution
$E$. This is shown in the following:

\begin{prop} Let us take the \syml\ $\widetilde{\CS}$ defined by $H_{2g+1}=1$
and
$H_i=0$ for $g+1\le i\le 2g$. Then, $\V{2g+1}(h_1,\dots,h_{2g+1})
\in \widetilde{\CS}$ for all $(h_1,\dots,h_{2g+1})\in \CM_{2g+1}$.
Moreover, the image $\widetilde\CQ$ of the previous map is transversal to $E$.
\end{prop}
{\bf Proof.}
By the
definition of $\widetilde{\CS}$, we have to show that $\frac12\Tr\left(
\V{2g+1}\right)^2=\la^{2g+2}+\sum_{i=0}^{g} H_i\la^i$. To this aim, we observe
that equation \rref{eq:e1} and the stationarity of $t_{2g+1}$ imply that
$\H{2g+1}(z)$ is an eigenvalue of $\V{2g+1}$. The other eigenvalue is given
by $\H{2g+1}(-z)$, because $\V{2g+1}$ depends only on $\la=z^2$.
Therefore, $\Tr\left(\V{2g+1}\right)^2=(\H{2g+1}(z))^2+(\H{2g+1}(-z))^2$
has the desired form.
Finally, referring to the parametrization \rref{paraS}, one can easily
prove that
the submanifold $p_{g}=0$ is transversal to the distribution $E$.
\endpf

Hence, 
the ${\mbox{KdV}}_{2g+1}$ phase space inherits from $\widetilde\CQ$ (and
from the quotient space $\widetilde\CN=\widetilde\CS/E$) a \bih\
structure. To compute this structure, it is convenient to use the formalism
discussed in \cite{cp}, whose aim is to avoid dealing with the explicit form of
the projection $\pi:\widetilde\CS\to\widetilde\CN$.

Now we want to show that the ${\mbox{KdV}}_{2g+1}$ flows
are indeed \bih\ with respect to the above Poisson pencil. The Lax
representation \rref{eq:Lax}
and the form \rref{forV} of the Lax pair suggest to consider
the vector fields on $\MA$ given by
\begin{equation}
\label{vfonM}
\dpt{X(\la)}{i}=\left[\left(\la^{i-g} X(\la)\right)_+,X(\la)\right]
\ ,\qquad i=-1,0,\dots,g-1.
\end{equation}
They are \ham\ \wrt\ the \poip\ on $\MA$, with \ham\ function given by
$X(\la)\mapsto \left(\la^{i-g}H(X(\la))\right)_+$, where
$H(X(\la))=\frac12\Tr X(\la)^2$. Furthermore, we can state

\begin{prop}
On the \bihm\ $\MA$ the function $H(\la)=\sum_{i=0}^{2g+1}H_i\la^i$
is a Casimir of the \poip.
The \bih\ \vefi\ $Y_{g-i}:=P_0 dH_{i-1}=P_1 dH_{i}$ 
has the Lax representation \rref{vfonM}.
\end{prop}
{\bf Proof.} One can easily see that $dH(\la)=\left[X(\la),
\la X(\la),\dots,\la^g X(\la)\right]^T$. Thus from \rref{eqcp0} and
\rref{eqcp1} it follows that $P_\la dH(\la)=0$. The \vefi s $Y_{i}$ are \ham\
\wrt\ $P_\la$, since $Y_{g-i}=P_1dH_i=P_\la d(\la^{-i}H(\la))_+$. Thus
the Lax representation \rref{vfonM} is a consequence of Proposition
\ref{bihtolax}.
\endpf

\begin{rem}
The previous proposition is a particular case of a general result \cite{RSTS},
stating that Ad-invariant polynomial functions on a Lie algebra $\alg$ give rise
to Casimir of the \poip\ on $\MA$.
\end{rem} 
In order to obtain the \kdv{2g+1}\ system from the \bih\ \vefi s \rref{vfonM},
we remark that, from general results of the \bih\ theory:
\begin{enumerate}
\item The functions $H_i$ are invariant along the distribution $E$, and
therefore can be projected on the quotient $\widetilde{\CN}$.
\item The \vefi s \rref{vfonM} are tangent to $\widetilde{\CS}$
and project onto $\widetilde{\CN}$.
\item Their projections are the \bih\ \vefi s associated with the projected
functions.
\end{enumerate}
We observe from \rref{vfonM} that the \vefi\ $Y_{-1}=P_0dH_g$ is tangent to the
distribution $E$. This means that the function $H_g$, on the quotient
$\widetilde\CN$, is a Casimir of the reduction of $P_0$. Hence, the polynomial
$H_0+H_1\la+\dots+H_g\la^g$ is a Casimir of the reduced \poip. The other \vefi s
in \rref{vfonM} project on the stationary reductions of KdV, as shown in
\begin{prop}
The projections on $\widetilde{\CN}$ of the \vefi s \rref{vfonM}, for
$i=0,\dots,g-1$, coincide
with the ${\mbox{KdV}}_{2g+1}$ systems.
\end{prop}
{\bf Proof.}
The right place to compare the two \hie ies of \vefi s is the
transversal submanifold $\widetilde{\CQ}$ defined as the image of the map
$\V{2g+1}:\CM_{2g+1}\to\widetilde{\CS}$. Hence, we must project the Lax
equations \rref{vfonM} on $T\widetilde{\CQ}$ along $E$. This leads
to
\begin{equation}
\label{decomp}
\left[\left(\la^{i-g} X(\la)\right)_+,X(\la)\right]-\alpha
\left[A,X(\la)\right]\ ,
\end{equation}
where $\alpha$ is fixed by the condition that this vector be tangent to
$\widetilde{\CQ}$. This means that the entry $(1,1)$ of the coefficient of
$\la^g$ in \rref{decomp} must be zero. If we write
$\left[\left(\la^{i-g} X(\la)\right)_+,X(\la)\right]
=\left[X(\la),\left(\la^{i-g} X(\la)\right)_-\right]$, then we obtain that
$\alpha$ is the entry $(1,2)$ of the $\la^{g-i-1}$--coefficient of $X(\la)$,
so that equation \rref{forV} concludes the proof.
\endpf

Therefore, we have shown that the \kdv{2g+1}\ flows are \bih.
Moreover, they are associated with a Casimir of the \poip, having
a polynomial dependence on $\la$. Since this is a particular
instance of a general theory developed by Gel'fand and
Zakharevich \cite{GZ93}, we will say that the \vefi s of \kdv{2g+1}\
are GZ systems. The next section is devoted to the separability
of such systems.

\section{Separability of \bih\ Systems}
\label{sec:sepbih}

We have just seen that the stationary reductions of KdV are
examples of GZ systems. In this section we show that the \bih\  structure
of such systems allows one to solve them by separation of variables.
Under special
circumstances, separability of GZ systems was proven in~\cite{Bl98, mt97}.
We refer to~\cite{fmp2}
for complete proofs and a more detailed discussion.

Let  $\CM$ be a $(2n+1)$-dimensional manifold endowed with a pencil $P_\la=
P_1-\la P_0$ of Poisson tensors. We suppose that the rank of $P_\la$ is
generically $2n$, so that (locally) there exists a
polynomial Casimir function $H(\la)=\sum_{i=0}^n H_i\la^i$ of $P_\la$
(see \cite{GZ93}).
The associated GZ systems $P_0 dH_i=P_1 dH_{i+1}$ are obviously
tangent to the \symls\ of $P_0$, and give rise to Liouville integrable
systems.
Since $H_n$ is a Casimir of $P_0$, such leaves are the
level surface of $H_n$. Let us denote by
$\omega$ the symplectic
form given by the restriction of $P_0$ to a (fixed) \syml; if
$X_f:=P_0 df$, where $f$ is any function on $\CM$, then
\[
\omega(X_f,X_g)=\{f,g\}_0\ .
\]
In order to exploit the existence of the other Poisson bracket,
we make an additional assumption. We suppose that there exists
a \vefi\ $Z$ on $\CM$ such that
\begin{enumerate}
\item It is transversal to the \symls\ of $P_0$;
\item The functions invariant along $Z$ form a {\em Poisson subalgebra}
\wrt\ the bracket $\parpu_\la$ associated with the Poisson pencil.
\end{enumerate}
The second condition means that the \bih\ structure can be projected
on the quotient space of the integral leaves of $Z$. The first one
tells us that such quotient can be identified with a \syml\ $\CS_c$
of $P_0$, which is therefore a \bihm.
Moreover, we can define on $\CS_c$ a {\em Nijenhuis tensor\/} $N$
as
\[
\omega(X_f,N X_g)=\{f,g\}_1\ ,
\]
where $f$ and $g$ are functions on $\CM$ invariant along $Z$.
Thus $\CS_c$ is said to be a Poisson--Nijenhuis (PN)
manifold (see~\cite{KM} and references cited therein).

The \vefi\ $Z$ allows us to use the \poip\ to construct variables
of separation for the GZ systems on $\CS_c$. Hence in this case
the \poip\ not only provides us with a commuting family of
\vefi s, but also gives coordinates 
for which the corresponding
equations of motion can be solved by separation of variables.
Indeed, one can show that the Nijenhuis tensor $N$
has $n$ functionally independent eigenvalues $(\la_1,\dots,\la_n)$.
Then (see, e.g., \cite{Ma90}), 
there exist $n$ complementary coordinates
$(\mu_1,\dots,\mu_n)$ on $\CS_c$ such that $\omega$ takes the
canonical form $\omega=\sum_{i=1}^n d\la_i\wedge d\mu_i$
and the adjoint $N^*$ of $N$ takes the diagonal form
\[
N^* d\la_j=\la_j d\la_j,\qquad N^* d\mu_j=\la_j d \mu_j\ .
\]
Such coordinates are called {\em Darboux--Nijenhuis (DN) coordinates\/}.
They are the separating coordinates for the GZ systems. Indeed, let
us normalize $Z$ in such a way that $Z(H_n)=1$. Then
the differentials of the restrictions ${\widehat H}_i$ of the \ham s
$H_i$ to $\CS_c$ generate a subspace which is invariant \wrt\ $N^*$. 
More precisely, we have that
\begin{equation}
\label{lenred}
\left[\begin{array}{c} {N}^* d{\widehat H}_0\\
                 \vdots \\
                 \vdots \\
                 \vdots \\
                 {N}^* d{\widehat H}_{n-1}\end{array}\right]
                 =
\left[\begin{array}{ccccc}
0 & 0 & \cdots & 0 & c_0 \\
1 & 0 & \cdots &        & c_1\\
0 & 1 & \cdot &         & c_2\\
\vdots &      & \cdot &  & \vdots\\
0 & 0 & \cdots & 1 & c_{n-1}
\end{array}\right]
\left[\begin{array}{c} d{\widehat H}_0\\
                 \vdots \\
                 \vdots \\
                 \vdots \\
                 d{\widehat H}_{n-1}\end{array}\right]\ ,
\end{equation}
where $c_i=-Z(H_i)$. This implies also that
\begin{equation}
\mbox{minimal polynomial of $N$}= \la^n-\sum_{i=0}^{n-1}c_i\la^i=Z(H(\la))\ .
\end{equation}
Moreover, one can check that the Frobenius matrix $\FF$ defined by
\rref{lenred} satisfies the condition
\begin{equation}
\label{sepacon}
{N}^* d\FF=\FF\, d\FF\ ,
\end{equation}
where $d\FF$ is the matrix whose entries are the
differentials of the entries of $\FF$, and, on the left--hand side,
$N^*$ acts separately on each entry. Conditions \rref{sepacon}
and \rref{lenred}
imply that the Hamilton-Jacobi equations for the ${\widehat H}_i$
are (collectively) separable in the DN coordinates. In fact, the
(transpose of the) Vandermonde matrix constructed with the $\la_j$
diagonalizes $\FF$,
and applied to $[{\widehat H}_0,\cdots,{\widehat H}_{n-1}]^T$
gives a {\em St\"ackel vector\/}, in the sense that
its $j$--th component depends only
on $(\la_j,\mu_j)$.

As far as the {\em explicit construction\/} of the DN coordinates
is concerned, we have seen that the $\lambda_j$ are the roots of the
minimal polynomial $Z(H(\la))$
of $N$. On the contrary, the coordinates $\mu_j$
must be computed  (in general)  by  a method involving  quadratures.
However, in the case at hand there is a recipe that is
particularly useful in the applications. Let us consider the
\ham\ \vefi\ $Y$ on $\CS_c$, associated with
$\frac12 \Tr N=\sum_{i=1}^n\la_i$ by the symplectic form $\omega$.
If the (restriction of the) Casimir $\widehat H(\la)$ satisfies
the condition $Y^r (\widehat H(\la))=0$ for some $r$, then
the coordinates
\[
\mu_j=\frac{Y^{r-2}(\widehat H(\la_j))}{Y^{r-1}(\widehat H(\la_j))}\
\]
form with the $\la_j$ a set of DN coordinates.
Hence in this case the \bih\ structure provides us with a
method to algebraically construct the separation variables. 

\section{Separability of the Stationary Reductions}
\label{sec:sepsta}

In this section we will show that the \kdv{2g+1}\ system
belongs to the class of separable GZ systems discussed above.
It is convenient to show that the conditions on the \vefi s
$Z$ and $Y$ are fulfilled on the ``big'' \bihm\ $\MA$ and then to
reduce everything.

Regarding the transversal \vefi\ $Z$, we introduce on $\MA$ the
\vefi\ $Z^\MA$ defined as
\begin{equation}
\label{ZonM}
{\dot X}_0=A,\qquad {\dot X}_i=0\ \mbox{ for all $i=1,\dots,g$.}
\end{equation}
It is tangent to the \symls\ of the \poit\ $P_0$ given by \rref{eqcp0}
(since it is easily seen to belong to its image) and it can be
projected on the quotient space ${\widetilde\CN}={\widetilde\CS}/E$
(since it commutes 
 with the generator \rref{genD} of the distribution
$E$). 
Using again the form \rref{eqcp0} and
\rref{eqcp1} of the \poit s on $\MA$, one can check that the functions invariant
along $Z^\MA$ form a Poisson subalgebra \wrt\ $P_\la$. This property
is trivially conserved after the reduction on ${\widetilde\CN}$.
Finally, we have to show that the reduced \vefi\ is transversal to
the \symls\ of (the reduction of) $P_0$. This follows from the fact
that, at the points where $\Tr(X_g A)=1$,
\[
L_{Z^\MA}(\textstyle{\frac12}\Tr X(\la)^2)=\Tr\left(X(\la)A\right)
=\la^g+\cdots\ ,
\]
so that $L_{Z^\MA} H_g=1$. This also shows that the reduction
of $Z^\MA$ has the right normalization.

Thus, we have shown that on $\widetilde\CN$, which is diffeomorphic to
the phase space of \kdv{2g+1}, there exists a \vefi\ that satisfies the
hypotheses of
the previous section. Therefore, the stationary reductions of
KdV can be solved by separation of variables in the DN
coordinates. We are left with the problem of finding explicitly
these coordinates.

To this aim, we introduce the \vefi\ $Y^\MA$ defined on $\MA$ as
\begin{equation}
\label{YonM}
{\dot X}_0=[A,X_g],\qquad {\dot X}_i=0\ \mbox{ for all $i=1,\dots,g$.}
\end{equation}
It is also tangent to the \symls\ of the \poit\ $P_0$
and can be projected on the quotient space ${\widetilde\CN}$.
Moreover,
\begin{enumerate}
\item $Y^\MA$ is (up to a sign) the \ham\
\vefi\ associated by means of $P_0$ with the Lie derivative along
$Z^\MA$ of the coefficient $H_{g-1}$ of $\frac12 \Tr X(\la)^2$;
\item We have that $L^2_{Y^\MA}(H(\la))=2\left(\Tr(A X_g)\right)^2$.
\end{enumerate}
The first assertion can be checked after noticing that
$Z^\MA(H_{g-1})=\Tr(X_{g-1}A)$, and that the differential
of this function is $\left[0,\dots,0,A,0\right]^T$. The second assertion
simply follows from the fact that $H(\la)=\frac12\Tr X(\la)^2$.

The same properties hold also on $\widetilde\CN$: If $Z$ and $Y$ are the
reductions of
$Z^\MA$ and $Y^\MA$, we have that $Y$ is the \ham\ \vefi\
associated with $-Z(H_{g-1})$ by the reduction of $P_0$. Furthermore,
we have that $L^2_Y(H(\la))=2$, since $\Tr(A X_g)=1$ at the points of
$\widetilde\CS$.
Let us now restrict ourselves 
to a \syml\ of the reduction of $P_0$. Since
$-Z(H_{g-1})=c_{g-1}=\frac12 \Tr N$, the \vefi\ $Y$ can be
used to construct the $\mu_j$ coordinates according to the
recipe given at the end of the previous section. The conclusion is:
\begin{enumerate}
\item The $\la_j$ are the roots of the polynomial
$Z(H(\la))=\Tr(A X(\la))=\Tr(A\V{2g+1})$,
that is, the entry $(1,2)$ of the matrix $\V{2g+1}$;
\item The $\mu_j$ are given by $\mu_j=f(\la_j)$, where
\[
f(\la)=\frac{Y(H(\la))}{Y^2(H(\la))}=\frac12 \Tr\left(X(\la)[A,X_g]\right)
=\mbox{entry $(2,2)$ of $\V{2g+1}$}.
\]
\end{enumerate} 
We remark that 
our general theory of separability gives, in this particular case,
the same construction of the variables of separation 
holding for systems admitting Lax with parameter formulation 
(see, e.g., \cite{AHH93,Sk}). In fact, writing the Lax matrix
\rref{pqtoV} as 
\[
\V{2g+1}=\left[
\begin{array}{cc}
V_g(\la) & U_g(\la)\\
W_g(\la) & -V_g(\la)
\end{array}\right]\ ,
\]
the equation of the associated spectral curve $C$ is 
\[
\mu^2=U_g(\la)W_g(\la)+V_g(\la)^2\ .
\]
Since $U_g(\la_j)=0$ and $\mu_j=-V_g(\la_j)$, we see that $g$ points 
$(\la_j,\mu_j)$ lie on $C$.

We close this section with a description, from our point of view, of the example
of \kdv5\ we started with in Section \ref{sec:preview}. We consider
the \bihm
\[
\MA=\left\{A\la^3+X_2\la^2+X_1\la+X_0\mid X_i=\mat2{p_i}{r_i}{q_i}{-p_i}
\right\}\ ,
\]
whose \poit s are given by \rref{eqcp0} and \rref{eqcp1}. The reduction process
described in Section \ref{sec:bihlax} allows us to pass to the transversal
submanifold
\[
A\la^3+\mat2{0}{1}{q_2}{0}\la^2+\mat2{p_1}{-q_2}{q_1}{-p_1}\la+
\mat2{p_0}{{q_2}^2-q_1}{q_0}{-p_0}\ ,
\]
which is diffeomorphic to the phase space $\CM_5$ of \kdv5. The correspondence
is given through the Lax matrix $\V5$ displayed in Section \ref{sec:lax}. 
The resulting change of variables is explicitly given by
\[
\begin{array}{c}
h_1=q_2\ ,\qquad h_2=-p_1\ ,\qquad h_3=q_1\ ,\qquad
h_4=-p_0-p_1 q_2\ ,\\
h_5=q_1 q_2+\frac12 {p_1}^2-\frac12 {q_2}^3+\frac12 q_0\ .
\end{array}
\]
The \poip\ on $\CM_5$ turns out to be
\begin{equation}
\label{redpoi5}
P_\la=\left [\begin {array}{ccccc}
0&-1&0&-h_1+\la&-h_2\\
1&0&2h_1-\la&h_2& h_3+\frac12{h_1}^{2} -2h_1\la\\
0&-2h_1+\la &0& -h_3-{h_1}^{2}+ 2h_1\la& -h_4-h_1h_2 \\
h_1-\la & -h_2& * &0&
\begin{array}{c} -h_5+3h_1h_3-\frac12{h_2}^{2}-{h_1}^{3}- \vspace{-2truemm}\\
(2h_3+{h_1}^{2})\la\end{array} \\
*&*&*&*&0
\end {array}\right ]
\end{equation}
Its Casimir $H(\la)=H_0+H_1\la+H_2\la^2$ can be computed with the trace of the
square of the Lax matrix:
\[
\begin{array}{l}
H_0=h_3{h_2}^{2}-2h_3h_5+{h_1}^{5}+2h_1{h_3}^{2}-2h_1h_2h_4-
3{h_1}^{3}h_3+2{h_1}^{2}h_5+{h_4}^{2}\\
H_1=2h_2h_4-2h_1h_5+3{h_1}^{2}h_3-h_1{h_2}^{2}-{h_3}^{2}-{h_1}^{4}\\
H_2=2{h_1}^{3}-4h_1h_3+2h_5
\end{array}
\]
The two \vefi s of the \kdv5\ \hie y are given by \rref{eqkdv5}. The \symls\ of
$P_0$ are the level surfaces of $H_2$. The \vefi\ $Z^\MA$ is
$\del/\del q_0$, while its projection $Z$ on $\CM_5$ is
$(1/2)\del / \del h_5$. 
On the \syml\ $\CS_c$ defined by $H_2=c$
we can use $(h_1,h_2,h_3,h_4)$ as global coordinates, and the corresponding
\poip\ is simply obtained by deleting the last row and the last column in
\rref{redpoi5}. The minimal polynomial of the Nijenhuis tensor on $\CS_c$ is
\[
Z(H(\la))=\la^2-h_1\la+h_1^2-h_3\ ,
\]
and $\la_1$, $\la_2$ are its roots.
To find $\mu_1$ and $\mu_2$ we have to use
\[
Y^\MA=-r_2\frac{\del}{\del p_0}+ 2p_2\frac{\del}{\del q_0}\ ,
\]
whose reduction on $\CM_5$ is $Y={\del}/{\del h_4}$.
Since $Y(H(\la))=2h_2\la+2h_4-2h_1h_2$ and $Y^2(H(\la))=2$, 
the coordinates $\mu_1$ and $\mu_2$ are
the values of the polynomial
\[
h_2\la+h_4-h_1h_2
\]
for $\la=\la_1,\la_2$. In order to check that the DN coordinates are separation
variables for the restrictions ${\widehat H}_0$ and ${\widehat H}_1$ of the
Hamiltonians to $\CS_c$, we simply have to compute
\[
\mat2{1}{\la_1}{1}{\la_2}\vec2{{\widehat H}_0}{{\widehat H}_1}
=\vec2{-c{\la_1}^2+{\mu_1}^2-{\la_1}^5}{-c{\la_2}^2+{\mu_2}^2-{\la_2}^5}\ ,
\]
from which the form of the spectral curve can also be seen.

\section{Final Remarks}
\label{sec:end}

1. The results outlined in Section \ref{sec:sepbih} are proved in \cite{fmp2}
for a class of \bihm s whose rank is not maximal. This means that our approach
to the stationary reductions of KdV
can be directly generalized to the stationary reductions of
the Gel'fand--Dickey hierarchies.
A step in this direction has already been taken in~\cite{fmt}, whose
results should be compared with those of~\cite{FH,Pr87}. We will treat this
problem in a future publication.

\noindent
2. The separation variables provided by the \bih\ method coincide, in the KdV
case, with the ones obtained by algebro-geometric constructions.
It would be interesting to compare in more general cases these two methods.
A first result has been obtained in \cite{Har}, where the ``spectral Darboux
coordinates'' of \cite{AHH93} are shown to be DN coordinates for a suitable pair
of compatible \poib s.

\noindent
3. Another Marsden-Ratiu reduction of the manifold $\MA$ of Section
\ref{sec:bihlax} has been performed in \cite{PeVa97}, for an arbitrary simple
Lie algebra $\alg$. That reduction leads to a bigger quotient space, and allows
one to reduce all the multi--\ham\ structure of $\MA$ and to obtain, in the case
$\alg=\fraksl(2)$, the Mumford systems \cite{Mum85}. A further \restr\ to the
level surface of some Casimirs gives the same reduced phase space obtained in
Section \ref{sec:bihstr}, where only two \poib s survive.

\subsection*{Acknowledgments}
J.P.Z. and M.P. were partially supported by FAPERJ 
through grant E-26/170.501/99-APV. J.P.Z is grateful to SISSA for its 
hospitality.
We thank G. Tondo for useful discussions at the early
stages of this work.
G.F. wishes to thank B. Dubrovin for useful discussions and remarks.
M.P. is grateful to IMPA and SISSA for their hospitality.

\end{document}